\documentstyle[prd,aps]{revtex}

\newcommand{\bea}{\begin{eqnarray}}
\newcommand{\eea}{\end{eqnarray}}

\begin{document}
\draft
\twocolumn[\hsize\textwidth\columnwidth\hsize\csname
@twocolumnfalse\endcsname

\title{Relativistic-Newtonian correspondence of the zero-pressure but \\
       weakly nonlinear cosmology}

\author{Hyerim Noh${}^{(a)}$ and Jai-chan Hwang${}^{(b)}$}
\address{${}^{(a)}$ Korea Astronomy and Space Science Institute,
                    Daejon, Korea \\
         ${}^{(b)}$ Department of Astronomy and Atmospheric Sciences,
                    Kyungpook National University, Taegu, Korea \\
         E-mails: ${}^{(a)}$ hr@kasi.re.kr, ${}^{(b)}$ jchan@knu.ac.kr
         }
\date{\today}
\maketitle

\begin{abstract}

It is well known that couplings occur among the scalar-, vector-,
and tensor-type perturbations of Friedmann world model in the second
perturbational order. Here, we {\it prove} that, except for the
gravitational wave contribution, the relativistic zero-pressure
irrotational fluid perturbed to second order in a flat Friedmann
background {\it coincides exactly} with the Newtonian result. Since
we include the cosmological constant, our results are relevant to
currently favoured cosmology. As we prove that the Newtonian
hydrodynamic equations are valid in {\it all} cosmological scales to
the second order, our result has an important practical implication
that one can now use the large-scale Newtonian numerical simulation
more reliably even as the simulation scale approaches and even goes
beyond the horizon. That is, our discovery shows that, in the
zero-pressure case, except for the gravitational wave contribution,
there are no relativistic correction terms even near and beyond the
horizon to the second-order perturbation.

\end{abstract}
\noindent

\vskip2pc]

%
%
\section{Introduction}

Historically, the first proper cosmological analysis appeared only
after the advent of Einstein's gravity theory in 1917
\cite{Einstein-1917}: the only known preceding cosmologically
relevant discussions can be found in Newton's correspondences to
Bentley in 1692 \cite{Newton-1692}. The expanding world model and
its linear structures were first studied in the context of
Einstein's gravity in the classic studies by Friedmann in 1922
\cite{Friedmann-1922} and Lifshitz in 1946 \cite{Lifshitz-1946},
respectively. In an interesting sequence, the much simpler and, in
hindsight, more intuitive Newtonian studies followed later by Milne
in 1934 \cite{Milne-1934} and Bonnor in 1957 \cite{Bonnor-1957},
respectively. According to Ellis ``It is curious that it took so
long for these dynamic models to be discovered after the (more
complex) general relativity models were known'' \cite{Ellis-1989}.
This is particularly so, because in the case without pressure the
Newtonian results {\it coincide exactly} with the previously derived
relativistic ones for both the background world model and its first
(linear) order perturbations. It would be fair to point out,
however, that the ordinarily known Newtonian cosmology (both for the
Friedmann background and its linear perturbations) is not purely
based on Newton's gravity, but is a guided one by Einstein's theory
\cite{Layzer-1954}.  The zero-pressure system with the cosmological
constant describes the current stage of our universe and its large
scale structures in the linear stage remarkably well.

Here we show that such a relativistic-Newtonian correspondence
continues even to the weakly nonlinear order. As the observed
large-scale structures show weakly nonlinear processes our
relativistic result has theoretical as well as practical
significances to interprete and analyse such structures properly in
relativistic level. As a consequence, our result implies that even
to the weakly nonlinear (second perturbational) order the well known
Newtonian equations can be used in all cosmological scales including
the super-horizon scale.

The known equations are
\bea
   \frac{\dot a^2}{a^2}
       = \frac{8 \pi G}{3} \varrho
       - \frac{\rm const.}{a^2}
       +\frac{\Lambda c^2}{3},
   \label{BG}
\eea
with $\varrho \propto a^{-3}$ for the background
\cite{Friedmann-1922,Milne-1934}, and
\bea
   \ddot \delta + 2 \frac{\dot a}{a} \dot \delta - 4 \pi G \varrho \delta
       = 0,
   \label{pert}
\eea for the linear order perturbations
\cite{Lifshitz-1946,Bonnor-1957}. The variable $a(t)$ is the scale
factor, and $\delta \equiv {\delta \varrho / \varrho}$ with
$\varrho$ and $\delta \varrho$ the background and perturbed parts of
the density field, respectively; $\Lambda$ is the cosmological
constant. The ``{\rm const.}'' part is interpreted as the spatial
curvature in Einstein's gravity \cite{Friedmann-1922,Robertson-1929}
and the total energy in the Newton's gravity \cite{Milne-1934}.
Although eq. (\ref{pert}) is also valid with general spatial
curvature the relativistic-Newtonian correspondence is somewhat
ambiguous in the case with curvature \cite{HN-Newtonian-1999}.
Therefore, in the following we consider the flat background only.
Equation (\ref{pert}) is valid even in the presence of the
cosmological constant $\Lambda$. We will include $\Lambda$ term in
our analyses of the weakly nonlinear stage. The above two equations
with vanishing spatial curvature describe remarkably well the
current expanding stage of our universe and its large-scale
structures which are believed to be in the linear stage. In the
small scale, however, the structures are apparently in nonlinear
stage, and even in the large scale weakly nonlinear study is needed.
Up until now, such a weakly nonlinear stage has been studied based
on Newton's gravity only.

The case with non-vanishing pressure {\it cannot} be handled in the
Newtonian context, especially for the perturbation. In this work, we
will show an additional continuation of relativistic-Newtonian
correspondences in the zero-pressure medium by showing that, except
for the gravitational wave contribution, the relativistic
second-order perturbation is described by the {\it same} set of
equations known in the Newtonian system. That is, except for the
coupling with the gravitational waves, the Newtonian equations {\it
coincide} exactly with the relativistic ones even to the second
order in perturbations.

{}For relativistic perturbations, due to the covariance of field
equations we have the freedom to fix the spacetime coordinates by
choosing some of the metric or energy-momentum variables at our
disposal: this is often called the gauge choice. The
relativistic-Newtonian correspondence to the linear order was made
by properly arranging the equations using various gauge-invariant
variables \cite{Bardeen-1980,HN-Newtonian-1999}. In the relativistic
case $\delta$ in eq. (\ref{pert}) is in fact $\delta$ in the
temporal comoving gauge condition which also implies the temporal
synchronous gauge in our zero-pressure case
\cite{Lifshitz-1946,Nariai-1969,Bardeen-1980}. In this work we
extend such correspondences to the second order.

%
%
\section{Fully nonlinear equations}

We may start from the completely nonlinear and covariant equations;
for a complete set of the covariant ($1+3$) equations see
\cite{covariant}. We consider a zero-pressure fluid with vanishing
isotropic pressure and anisotropic stress, thus $\tilde p \equiv 0
\equiv \tilde \pi_{ab}$. In the energy frame we take $\tilde q_a
\equiv 0$. Tildes indicate the covariant quantities, and the Greek
and Latin indices indicate the space and spacetime indices,
respectively. The momentum conservation gives vanishing acceleration
vector $\tilde a_a$ to all orders; see eq. (27) of \cite{NL}. The
energy conservation and the Raychaudhury equation ($\tilde
G^\alpha_\alpha - \tilde G^0_0$ part of Einstein's equation) give
\bea
   \tilde {\dot {\tilde \mu}} + \tilde \mu \tilde \theta
   &=& 0,
   \label{covariant-eq1} \\
   \tilde {\dot {\tilde \theta}} + \frac{1}{3} \tilde \theta^2
       + \tilde \sigma^{ab} \tilde \sigma_{ab}
       - \tilde \omega^{ab} \tilde \omega_{ab}
       + 4 \pi G \tilde \mu - \Lambda
   &=& 0,
   \label{covariant-eq2}
\eea see eqs. (26,28) of \cite{NL}; $\tilde \theta \equiv \tilde
u^a_{\;\; ;a}$ is an expansion scalar with $\tilde u_a$ a fluid
four-vector, $\tilde \sigma_{ab}$ and $\tilde \omega_{ab}$ are the
shear and the rotation tensors, respectively; if $\tilde u_\alpha =
0$ we have no rotation of the fluid four-vector $\tilde u_a$. We set
$c \equiv 1$. An overdot with tilde is a covariant derivative along
the $\tilde u_a$ vector, e.g., $\tilde {\dot {\tilde \mu}} \equiv
\tilde \mu_{,a} \tilde u^a$. By combining these equations we have
\bea
   & & \left( \frac{\tilde {\dot {\tilde \mu}}}{\tilde \mu}
       \right)^{\tilde \cdot}
       - \frac{1}{3}
       \left( \frac{\tilde {\dot {\tilde \mu}}}{\tilde \mu} \right)^2
       - \tilde \sigma^{ab} \tilde \sigma_{ab}
       + \tilde \omega^{ab} \tilde \omega_{ab}
       - 4 \pi G \tilde \mu
       + \Lambda
       = 0.
   \nonumber \\
   \label{covariant-eq3}
\eea Equations (\ref{covariant-eq1}-\ref{covariant-eq3}) are valid
to all orders, i.e., these equations are fully nonlinear and still
covariant.

We consider the {\it scalar-} and {\it tensor-type} perturbations in
the {\it flat} Friedmann background without pressure. We ignore the
vector-type perturbation; the vector-type perturbation (rotation) is
supposed to be not important because it always decays in the
expanding phase even to the second order, see \S VII.E of \cite{NL}.
We will take the {\it comoving} gauge, and by ignoring the
vector-type perturbations we have no rotation. In this case the
four-vector becomes $\tilde u_\alpha = 0$, thus coincides with the
normal four-vector $\tilde n_a$. We lose no generality by imposing
the gauge condition. In our case the energy-momentum tensor becomes
$\tilde T^0_0 = - \tilde \mu$, and $\tilde T^0_\alpha = 0 = \tilde
T^\alpha_\beta$ where $\tilde \mu$ is the energy density. We
emphasize that as our comoving gauge condition fixes the gauge
degree of freedom completely, all variables in this gauge condition
are equivalently gauge invariant to the second order: this is in the
sense that each of the variables has a unique corresponding
gauge-invariant combination, see \cite{NL}.

In \cite{NL} the equations are presented in the normal frame $\tilde
n_a$ with $\tilde n_\alpha \equiv 0$. The fluid four-vector $\tilde
u_a$ in general differs from the normal four-vector $\tilde n_a$.
{\it Only} in the comoving gauge without rotation the two frames
coincide. Since the fluid quantities are defined in the fluid
($\tilde u_a$) frame, the zero-pressure condition should be imposed
in $\tilde u_a$ frame. Thus, for the fluid quantities defined in the
normal frame the physical zero-pressure condition implies vanishing
pressures (both isotropic and anisotropic) only in the comoving
gauge without rotation. In this normal frame, the gauge
transformation to the second order causes pressure terms to appear
in other gauges, see \cite{NL}. In the energy frame, which takes
vanishing flux $\tilde q_a \equiv 0$ as the frame condition, the
comoving gauge condition takes $\tilde u_\alpha \equiv 0$ for the
fluid four-vector; here, we ignore the vector-type perturbation.
Since $\tilde u_\alpha = 0$ it coincides with the normal frame
vector. Now, in the normal frame, which takes $\tilde n_\alpha
\equiv 0$ as the frame condition, the comoving gauge condition
without rotation implies $\tilde q_a \equiv 0$. Thus, as long as we
take the comoving gauge without rotation, in either frame we have
$\tilde q_a \equiv 0$ and $\tilde u_\alpha = 0 = \tilde n_\alpha$;
i.e., the fluid four-vector coincides with the normal four-vector.

%
%
\section{Correspondence to the second order: a proof}

Now, we consider equations perturbed to the second order
in the metric and matter variables.
We introduce
\bea
   \tilde \mu \equiv \mu + \delta \mu, \quad
       \tilde \theta \equiv 3 \frac{\dot a}{a} + \delta \theta,
\eea
where $\mu$ and $\delta \mu$ are the background and perturbed energy density,
respectively, and $\delta \theta$ is the perturbed part of expansion scalar;
we set $\delta \equiv {\delta \mu / \mu}$.
We {\it identify} $\mu \equiv \varrho$ to the background order, and
\bea
   \delta \mu \equiv \delta \varrho, \quad
       \delta \theta \equiv \frac{1}{a} \nabla \cdot {\bf u},
   \label{identification}
\eea to both the linear- and second-order perturbations. Now, to the
second order, after some algebra using perturbed order quantities
presented in \cite{NL}, the perturbed parts of eqs.
(\ref{covariant-eq1},\ref{covariant-eq2}) give (see the Appendix)
\bea
   & & \dot \delta + \frac{1}{a} \nabla \cdot {\bf u}
       = - \frac{1}{a} \nabla \cdot \left( \delta {\bf u} \right),
   \label{dot-delta-eq} \\
   & & \frac{1}{a} \nabla \cdot \left( \dot {\bf u}
       + \frac{\dot a}{a} {\bf u} \right)
       + 4 \pi G \mu \delta
   \nonumber \\
   & & \qquad
       = - \frac{1}{a^2} \nabla \cdot
       \left( {\bf u} \cdot \nabla {\bf u} \right)
       - \dot C^{(t)\alpha\beta} \left( \frac{2}{a} \nabla_\alpha u_\beta
       + \dot C^{(t)}_{\alpha\beta} \right),
   \label{dot-delta-u-eq}
\eea where $C^{(t)}_{\alpha\beta}$ is the transverse and tracefree
tensor-type perturbation (the gravitational waves) introduced in eq.
(\ref{metric}); the indices of $C^{(t)}_{\alpha\beta}$ are raised
and lowered by $\delta_{\alpha\beta}$; its contribution in eq.
(\ref{dot-delta-u-eq}) comes from the shear term in eq.
(\ref{covariant-eq2}); see the Appendix. By combining these
equations we have \bea
   & & \ddot \delta + 2 \frac{\dot a}{a} \dot \delta - 4 \pi G \mu \delta
       = - \frac{1}{a^2} \frac{\partial}{\partial t}
       \left[ a \nabla \cdot \left( \delta {\bf u} \right) \right]
       + \frac{1}{a^2} \nabla \cdot \left( {\bf u} \cdot \nabla {\bf u} \right)
   \nonumber \\
   & & \qquad
       + \dot C^{(t)\alpha\beta} \left( \frac{2}{a} \nabla_\alpha u_\beta
       + \dot C^{(t)}_{\alpha\beta} \right),
   \label{ddot-delta-eq}
\eea which also follows from eq. (\ref{covariant-eq3}). Equations
(\ref{dot-delta-eq}-\ref{ddot-delta-eq}) are our extension of eq.
(\ref{pert}) to the second-order perturbations in Einstein's theory.
We will show that, except for the gravitational wave contribution,
exactly the same equations also follow from Newton's theory.

The presence of linear-order gravitational waves can generate the
second-order scalar-type perturbation by generating the shear terms.
The coupling between the scalar-type perturbation and the
gravitational waves to the second order was noticed in the original
study of the second-order perturbations by Tomita in 1967
\cite{Tomita-1967}.  Here, we note the behaviour of the
gravitational waves in the linear regime. To the linear order the
gravitational waves are described by the well known wave equation
\cite{Lifshitz-1946} \bea
   & & \ddot C^{(t)}_{\alpha\beta}
       + 3 {\dot a \over a} \dot C^{(t)}_{\alpha\beta}
       - {\Delta \over a^2} C^{(t)}_{\alpha\beta} = 0.
\eea In the super-horizon scale the non-transient mode of
$C^{(t)}_{\alpha\beta}$ remains constant, thus $\dot
C^{(t)}_{\alpha\beta} = 0$, and in the sub-horizon scale, the
oscillatory $C^{(t)}_{\alpha\beta}$ redshifts away, thus
$C^{(t)}_{\alpha\beta} \propto a^{-1}$. Notice that only time
derivatives of $C^{(t)}_{\alpha\beta}$ generate the scalar-type
perturbation. Thus, we anticipate that the contribution of
gravitational waves to the scalar-type perturbation is not
significant to the second order. The quadratic combinations of
linear-order scalar-type perturbation can also work as sources for
the gravitational waves to the second-order \cite{second-order}.

In the Newtonian context, the mass conservation, the momentum conservation, and
the Poisson's equation give \cite{Peebles-1980}
\bea
   \dot \delta + \frac{1}{a} \nabla \cdot {\bf u}
   &=& - \frac{1}{a} \nabla \cdot \left( \delta {\bf u} \right),
   \label{mass-conservation} \\
   \dot {\bf u} + \frac{\dot a}{a} {\bf u} + \frac{1}{a} \nabla \delta \Phi
   &=& - \frac{1}{a} {\bf u} \cdot \nabla {\bf u},
   \label{momentum-conservation} \\
   \frac{1}{a^2} \nabla^2 \delta \Phi
   &=& 4 \pi G \varrho \delta,
   \label{Poisson-eq}
\eea where $\delta \Phi$ is the perturbed gravitational potential.
We note that these equations are valid to fully nonlinear order.
Equation (\ref{dot-delta-eq}) follows from eq.
(\ref{mass-conservation}), and eq. (\ref{dot-delta-u-eq}) ignoring
the gravitational waves follows from eqs.
(\ref{momentum-conservation},\ref{Poisson-eq}). Thus, eq.
(\ref{ddot-delta-eq}) also naturally follows in Newton's theory
\cite{Peebles-1980}. In this Newtonian situation ${\bf u}$ is the
perturbed velocity and $\delta \equiv {\delta \varrho / \varrho}$.
This completes our proof of the relativistic-Newtonian
correspondence to the second order. Although we successfully
identified the relativistic density and velocity perturbation
variables we do {\it not} have a relativistic variable which
corresponds to the Newtonian potential $\delta \Phi$ to the second
order. This situation could be understood because the gravitational
potential introduced in Poisson's equation reveals the
action-at-a-distance nature and the static nature of Newton's
gravity theory compared with Einstein's gravity.

Notice that in the Newtonian context eqs.
(\ref{dot-delta-eq},\ref{dot-delta-u-eq}), thus eq.
(\ref{ddot-delta-eq}) as well, without the gravitational waves, are
valid to fully nonlinear order. This has an important implication
that any nonvanishing third and higher order correction terms in
relativistic context should be regarded as purely relativistic
effects \cite{third-order}. To our knowledge the Newtonian equations
in eqs. (\ref{mass-conservation}-\ref{Poisson-eq}) were first
presented by Peebles in \cite{Peebles-1980}.

%
%
\section{Discussions}

We have shown that to the second order, except for the gravitational
wave contribution, the zero-pressure relativistic cosmological
perturbation equations can be exactly identified with the known
equations in Newton's theory. As a consequence, to the second order,
we identified the correct relativistic variables which can be
interpreted as density $\delta \mu$ and velocity $\delta \theta$
perturbations in eq. (\ref{identification}), and we showed that to
that order the Newtonian hydrodynamic equations remain valid in all
cosmological scales including the super-horizon scale. Our results,
showing the equivalence to the second order of the zero-pressure
relativistic scalar-type perturbation and the Newtonian ones, may
not be entirely surprising considering the Birkhoff's theorem
\cite{Birkhoff-1923}; for cosmology related discussions of the
theorem, see \cite{Peebles-1980}. However, our results should not be
so obvious either, because our system lacks any spatial symmetry
contrary to the Birkhoff's theorem which is concerned with the
spherically symmetric system. It might be as well that our
relativistic results give relativistic correction terms appearing to
the second order which become important as we approach and go beyond
the horizon scale. Our results show that there are {\it no} such
correction terms appearing to the second order, and the
correspondence is {\it exact} to that order. A complementary result,
showing the relativistic-Newtonian correspondence in the Newtonian
limit of post-Newtonian approach ($\frac{v}{c}$-expansion with
$\frac{GM}{Rc^2} \sim \frac{v^2}{c^2} \ll 1$, thus valid far inside
the horizon), can be found in \cite{post-Newtonian}. In fact, the
Newtonian hydrodynamic equations naturally appear in the zeroth
post-Newtonian order of Einstein's gravity
\cite{Chandrasekhar-1965}; for the cosmological extension, see
\cite{PN}.

In a classic study of the cosmic microwave background radiation
anisotropy in 1967 Sachs and Wolfe have mentioned that ``the linear
perturbations are so surprisingly simple that a perturbation
analysis accurate to second order may be feasible using the methods
of Hawking (1966)'' \cite{SW-1967}. Our proof of the exact
relativistic-Newtonian correspondence to the second order could be
regarded as one of such accurate results anticipated in
\cite{SW-1967}. Indeed, in this work we used the method of Hawking
which is the covariant (1+3) equations \cite{Hawking-1966}; for
other proofs see \cite{second-order}.

As we consider a flat background the ordinary Fourier analysis can
be used to study the mode-couplings as in the Newtonian case
\cite{quasilinear}. Our equations include the cosmological constant,
thus compatible with current observations of the large-scale
structure and the cosmic microwave background radiation which favour
nearly flat Friedmann world model with non-vanishing $\Lambda$
\cite{observations}. Our result may also have the following
important practical cosmological implication. As we have proved that
the Newtonian hydrodynamic equations are valid in {\it all}
cosmological scales to the second order, our result has an important
comological implication that one can use the large-scale Newtonian
numerical simulation more reliably in the general relativistic
context even as the simulation scale approaches near (and goes
beyond) horizon scale.  The fluctuations near the horizon scale are
supposed to be linear or weakly nonlinear; otherwise, it is
difficult to imagine the presence of spatially homogeneous and
isotropic background world model which is the basic assumption and
the backbone of the modern cosmology.

Since the Newtonian system is exact to the second order in
nonlinearity, any non-vanishing third and higher order perturbation
terms in the relativistic analysis can be regarded as the pure
relativistic correction. Expanding the fully nonlinear equations in
(\ref{covariant-eq1}-\ref{covariant-eq3}) to third and higher order
will give the potential correction terms. {}For our recent work in
the third-order perturbations, see \cite{third-order}. In
\cite{third-order} we derive the nonvanishing third-order terms
which are purely relativistic correction terms. We also show that
these correction terms are independent of the horizon and are
smaller than the second-order Newtonian/relativistic terms by a
factor $10^{-5}$, thus negligible indeed.

%
%
\subsection*{Acknowledgments}

We thank Dirk Puetzfeld for helpful comments. We wish to thank Lev
Kofman and Ethan Vishniac for their comments on possible relation to
the Birkhoff's theorem. HN and JH were supported by grants No.
R04-2003-10004-0 and No. R02-2003-000-10051-0, respectively, from
the Basic Research Program of the Korea Science and Engineering
Foundation.

%
%
\section*{Appendix}

         Since eqs. (\ref{dot-delta-eq},\ref{dot-delta-u-eq})
         allow us to conclude about the
         relativistic-Newtonian correspondence, in the following
         we will derive these equations in some detail
         using the basic quantities presented in \cite{NL}.
         As the metric we take
         \bea
            & & ds^2 = - a^2 \left( 1 + 2 \alpha \right) d \eta^2
                - 2 a \chi_{,\alpha} d \eta d x^\alpha
            \nonumber \\
            & & \qquad
                + a^2 \left[ \left( 1 + 2 \varphi \right) \delta_{\alpha\beta}
                + 2 C^{(t)}_{\alpha\beta} \right] d x^\alpha d x^\beta,
            \label{metric}
         \eea
         which follows from our convention in eqs. (49,175,178) of \cite{NL}.
         Here, $\alpha$, $\chi$ and $\varphi$ are spacetime dependent
         perturbed-order variables; $C^{(t)}_{\alpha\beta}$ is a
         transverse-tracefree tensor-type perturbation variable.
         In our metric we took the spatial $C$-gauge which removes the spatial
         gauge modes completely, thus all the remaining
         variables we are using are spatially gauge-invariant,
         see \S VI.B.2 of \cite{NL}; to the linear order,
         our notation coincides with Bardeen's in \cite{Bardeen-1988}.
         We work in the temporal comoving gauge which takes $v=0$ in \cite{NL}.
         The momentum conservation gives $\tilde a_\alpha = 0$ which gives
         $\alpha = - \frac{1}{2 a^2} \chi^{,\alpha} \chi_{,\alpha}$,
         see eq. (69) of \cite{NL}.
         In our comoving gauge condition, the four-vector
         in eq. (53) of \cite{NL}, using eq. (175) in that paper, becomes
         \bea
            & & \tilde u_0 = - a, \quad
                \tilde u_\alpha = 0;
            \nonumber \\
            & &
                \tilde u^0 = \frac{1}{a}, \quad
                \tilde u^\alpha = \frac{1}{a^2} \chi^{,\beta}
                \left[ \left( 1 - 2 \varphi \right) \delta^\alpha_\beta
                - 2 C^{(t)\alpha}_{\;\;\;\;\;\beta} \right].
         \eea
         With these, we have
         \bea
            \tilde {\dot {\tilde \mu}}
            &=& \tilde \mu_{,0} \tilde u^0
                + \tilde \mu_{,\alpha} \tilde u^\alpha
                = \left[ \mu \left( 1 + \delta \right) \right]^\cdot
                + \frac{1}{a^2} \mu \delta_{,\alpha} \chi^{,\alpha},
            \nonumber \\
            \tilde {\dot {\tilde \theta}}
            &=& \left( 3 \frac{\dot a}{a} + \frac{1}{a} \nabla \cdot {\bf u}
                \right)^\cdot
                + \frac{1}{a^3}
                \left( \nabla \cdot {\bf u} \right)_{,\alpha} \chi^{,\alpha}.
         \eea
         Our $\delta \theta$ is the same as $- \kappa$ in \cite{NL}.
         Using eqs. (55,57,70) of \cite{NL} we can show
         \bea
            & & \tilde \sigma^{ab} \tilde \sigma_{ab}
                = \frac{1}{a^4} \left[ \chi^{,\alpha\beta} \chi_{,\alpha\beta}
                - \frac{1}{3} \left( \Delta \chi \right)^2 \right]
            \nonumber \\
            & & \qquad
                + \dot C^{(t)\alpha\beta}
                \left( \frac{2}{a^2} \chi_{,\alpha\beta}
                + \dot C^{(t)}_{\alpha\beta} \right).
         \eea
         Now, we have to relate $\chi$ to our notation.
         Apparently, we need $\chi$ only to the linear order.
         The $\tilde G^0_\alpha$-component of Einstein's equation in eq. (197) of
         \cite{NL} gives
         $\frac{\Delta}{a^2} \chi = - \kappa \equiv \delta \theta
         \equiv \frac{1}{a} \nabla \cdot {\bf u}$.
         As our ${\bf u}$ is of the potential type,
         i.e., of the form ${\bf u} \equiv u_{,\alpha}$,
         we have ${\bf u} = \frac{1}{a} \nabla \chi$
         to the linear order.
         Using these, eqs. (\ref{covariant-eq1},\ref{covariant-eq2}) give
         \bea
            & & \left( \frac{\dot \mu}{\mu} + 3 \frac{\dot a}{a} \right)
                \left( 1 + \delta \right)
                + \dot \delta
                + \frac{1}{a} \nabla \cdot {\bf u}
                = - \frac{1}{a} \nabla \cdot \left( \delta {\bf u} \right),
            \\
            & & 3 \frac{\ddot a}{a} + 4 \pi G \mu - \Lambda
                + \frac{1}{a} \nabla \cdot \left( \dot {\bf u}
                + \frac{\dot a}{a} {\bf u} \right)
                + 4 \pi G \mu \delta
            \nonumber \\
            & & \qquad
                = - \frac{1}{a^2}
                \nabla \left( {\bf u} \cdot \nabla {\bf u} \right)
                - \dot C^{(t)\alpha\beta} \left( \frac{2}{a} u_{\alpha,\beta}
                + \dot C^{(t)}_{\alpha\beta} \right).
         \eea
         To the background order, we have
         $\dot \mu + 3 (\dot a/a) \mu = 0$ and
         $3\ddot a/a + 4 \pi G \mu - \Lambda = 0$;
         after an integration we recover eq. (\ref{BG}) with $\Lambda$;
         in this case the ``{\rm const.}'' is an integration constant which
         can be interpreted as the spatial curvature.
         The perturbed parts give
         eqs. (\ref{dot-delta-eq},\ref{dot-delta-u-eq}).


\end{document}